\documentclass[11pt]{article}

\usepackage{cite}

\newcommand{\ben}{\begin{enumerate}}
\newcommand{\een}{\end{enumerate}}
\newcommand{\be}{\begin{equation}}
\newcommand{\ee}{\end{equation}}
\newcommand{\bse}{\begin{subequation}}
\newcommand{\ese}{\end{subequation}}
\newcommand{\bea}{\begin{eqnarray}}
\newcommand{\eea}{\end{eqnarray}}
\newcommand{\bc}{\begin{center}}
\newcommand{\ec}{\end{center}}

\def\upe#1{\mathop{}\limits^{\raise8pt\hbox{#1}}}
\def\upv#1{\mathop{\vert}\limits^{\raise4pt\hbox{#1}}}
\def\rigat#1{\mathop{\rightarrow}\limits^{#1}}
\def\downb#1{\mathop{\bullet}\limits_{#1}}

\begin{document}

\rightline{UNN-SCM-M-00-6}
\rightline{October 2000}
\rightline{revised April 2001}
\rightline{J. Phys. {\bf A34} (2001) L503-L509}
\vspace{1.7 cm}

\begin{center}

 \textsf{\LARGE Simple Applications of $q$-Bosons}

 \vspace{7mm}

{\textbf{Maia Angelova}$^{\dagger }$\textbf{, V.K.
Dobrev}$^{\dagger }$\footnote{Permanent address:
Institute of Nuclear Research and Nuclear Energy,
Bulgarian Academy of Sciences, Sofia, Bulgaria.}
\textbf{\ and A. Frank}$^{\ddagger }$

\vspace{5mm}

$^{\dagger }$\textit{School of Computing and Mathematics}\\
\textit{University of Northumbria, Newcastle upon Tyne, UK GB-NE1 8ST}\\[2mm]
$^{{}}\ddagger $\textit{Instituto de Ciencias Nucleares and Centro de
Ciencias F\'{i}sicas,}\\
\textit{UNAM, A.P. 70-543, Mexico, D.F., 04510 Mexico.}\\
}

\end{center}

\vspace{.8 cm}

\begin{abstract}
A deformation of the harmonic oscillator algebra associated with
the Morse potential and the $SU(2)$ algebra is derived using the
quantum analogue of the anharmonic oscillator. We use the
quantum oscillator algebra or $q$-boson algebra which is a
generalisation of the Heisenberg-Weyl algebra obtained by
introducing a deformation parameter $q$. Further, we present a
new algebraic realization of the $q$-bosons, for the case of $q$
being a root of unity, which corresponds to a periodic structure
described by a finite-dimensional representation. We show that
this structure represents the symmetry of a linear lattice with
periodic boundary conditions.

\end{abstract}


\vspace{1 cm}

\section{Introduction}

Algebraic models have been used very successfully in nuclear and
molecular physics and have led to new insights into the nature of
complex many body systems \cite{iac:interacting, frank:methods,
iac:algebraic}. The methods combine Lie algebraic techniques,
describing the interatomic interactions, with discrete symmetry
techniques associated with the global symmetry of the atoms and
molecules in complex compounds. In the framework of the algebraic
model \cite{frank:methods}, the anharmonic effects of the local
interactions are described by substituting the local harmonic
potentials by Morse-like potential. The Morse potential, which is
associated with the $SU(2)$ algebra, leads to a deformation of
the harmonic oscillator algebra.

In this paper we derive the above-mentioned deformation using the
quantum analogue of the anharmonic oscillator. We describe the
anharmonic vibrations as anharmonic $ q $-bosons. Their algebra,
known as quantum oscillator algebra\ $HW_q$\ or $q$-boson
algebra, has been introduced in
\cite{arik-coon:jmp,biedenharn:jphysa,macfarlane:jphysa}, and is
a generalisation of the Heisenberg-Weyl algebra obtained by
introducing a deformation parameter $q$\textit{.} A change of
parametrization leads naturally to Lie-algebraic approximations
of the general $q$-deformation. In particular, we present a new
algebraic realization of the $q$-boson algebra, for the case of
$q$ being a root of unity, which corresponds to a periodic
structure described by a finite-dimensional representation.
We show that this structure generates a group isomorphic to
the symmetry group of a linear lattice with periodic boundary conditions.

\section{\protect\bigskip Algebraic Model}

\bigskip In the algebraic model \cite{frank:methods}, the
one-dimensional Morse Hamiltonian is written in terms of the
\bigskip generators $\ $of $SU(2)$,
\begin{equation}
H_{M}=\frac{A}{4}\left(
\hat{\mathcal{N}}^{2}-4\hat{J}_{Z}^{2}\right) =\frac{
A}{2}(\hat{J}_{+}\hat{J}_{-}+\hat{J}_{-}\hat{J}_{+}-\hat{\mathcal{N}})
\end{equation}
where $A$ is a constant. The eigenstates,
$|\![\mathcal{N}],v\rangle $, correspond to the $\ U(2)\supset
SU(2)$ symmetry-adapted basis,$\;$where $
\mathcal{N}\;$is the total number of bosons fixed by the
potential shape, and $v$ is the number of quanta in the
oscillator, $v=1,2,\ldots ,\left[
\frac{\mathcal{N}}{2}\right] $.

The anharmonic effects are described by anharmonic boson
operators \cite {frank:methods},
\begin{equation}
\hat{b}=\frac{\hat{J}_{+}}{\sqrt{\mathcal{N}}},\;\;\;\hat{b}^{\dagger }=
\frac{\hat{J}_{-}}{\sqrt{\mathcal{N}}}\;,\;\;\;\hat{v}=\frac{\hat{\mathcal{N}
}}{2}-\hat{J}_{z}
\end{equation}
where $\hat{v}\;\;$is the Morse phonon operator with an
eigenvalue $v$. The operators satisfy the commutation relations,
\begin{equation}
\left[ \hat{b},\hat{v}\right] =\hat{b},\;\;\;\;\;\;\;\left[ \hat{b}
^{^{\dagger }},\hat{v}\right] =-\hat{b}^{^{\dagger
}},\;\;\;\;\left[\hat{b},
\hat{b}^{^{\dagger }}\right] =1-\frac{2\hat{v}}{\mathcal{N}} .
\label{anh}
\end{equation}
For an infinite potential depth, $\mathcal{N}\rightarrow \infty
$, \ \ $
\left[ \hat{b},\hat{b}^{^{\dagger }}\right] \rightarrow 1,$ \ giving the
usual boson commutation relations associated with the harmonic
oscillator.

The Morse Hamiltonian can be written in terms of \ the anharmonic
bosons $\ \hat{b}$ and $\hat{b}^{^{\dagger }}$,
\begin{equation}
H_{M}\sim \frac{1}{2}\left( \hat{b}\hat{b}^{^{\dagger
}}+\hat{b}^{^{\dagger }}\hat{b}\right)
\end{equation}
which corresponds to vibrational energies
\begin{equation} \label{vibr}
\varepsilon _{v}=\hbar \omega _{0}\left(
v+\frac{1}{2}-\frac{v^{2}}{\mathcal{N}}\right)
\end{equation}
where $\omega _{0}$ is the harmonic oscillator frequency. When
$\;\;\mathcal{ N}\rightarrow \infty $, the Morse potential cannot
be distinguished from the harmonic potential.

The algebraic model has been developed to analyse molecular
vibrational spectra \cite{frank:methods, iac:algebraic, lemus:chem,
frank:chem, frank:annals, perez:mol, oss:chem}.
It provides a systematic procedure for studying vibrational
excitations in a simple form by describing the stretching and
bending modes in a unified scheme based on $SU(2)$ algebra which
incorporates the anharmonicity at the local level.

\section{\protect\bigskip Anharmonic $q$-Bosons}

The anharmonic bosons of the previous section may be obtained as
an approximation of the so-called q-bosons
\cite{arik-coon:jmp,biedenharn:jphysa,macfarlane:jphysa}, which
enter the Heisen\-berg-Weyl $q$-algebra $HW_{q}$ \ given by the
following commutation relations:
\begin{equation}
\lbrack a,a^{\dagger }]=q^{\hat{n}}\ ,\quad \lbrack \hat{n},a]=-a\
,\quad\lbrack \hat{n},a^{\dagger }]=a^{\dagger } \label{qanh}
\end{equation}
where $q$ is in general a complex number. This number is called
deformation parameter since the boson commutation relations of
the harmonic oscillator may be recovered for the value $q=1$.

For the general analysis of the system (\ref{qanh}) it is useful
to know operators which are diagonalizable. Such an operator is
the Casimir operator which for $HW_{q}$ can be written in the form:
\begin{equation}
\mathcal{C}=aa^{\dagger }+a^{\dagger
}a-\frac{q^{{\hat{n}}+1}+q^{\hat{n}}-2}{q-1} .
\end{equation}
Checking that:
\begin{equation}
\lbrack \mathcal{C},a]=[\mathcal{C},a^{\dagger
}]=[\mathcal{C},\hat{n}]=0
\end{equation}
is straightforward. Thus, a possible Hamiltonian is:
\begin{equation}
H=\frac{1}{2}(aa^{\dagger }+a^{\dagger }a)=
\frac{1}{2}\mathcal{C}+\frac{1}{2} \frac{q^{\hat{n}
+1}+q^{\hat{n}}-2}{q-1} .\label{hamq}
\end{equation}

The anharmonic bosons (\ref{anh}) may obtained from the
$q$-bosons (\ref {qanh}) for real values of $q$ close to 1.
Namely, let $q<1$ and $p\equiv 1/(1-q)$, so that $\ q=1-1/p$ \
and
\begin{equation}
q^{\hat{n}}=(1-\frac{1}{p})^{\hat{n}}\,.
\end{equation}
The harmonic limit is recovered for $p\rightarrow \infty $ in
this parametrization.

Further, assuming that $1/p\ll 1$ and neglecting the terms of
order $1/p^{2}$
\ and higher, we obtain,
\begin{equation}
q^{\hat{n}}=1-\frac{\hat{n}}{p} ~~ . \label{qqanh}
\end{equation}
If we now substitute the approximation for $q^{\hat n}$ from eq.
(\ref{qqanh}) in the commutation relations (\ref{qanh}) and
identify the parameter $p$ with ${\cal N}/2$, $\hat n$ with $\hat
v$ and the creation and annihilation operators $a, a^\dagger$,
with $b, b^\dagger$, we recover the $SU(2)$ anharmonic relations
(\ref{anh}). We can now explain the meaning of this approximation. In a
sense, the form (\ref{anh}) of the $SU(2)$ commutation relations can be
considered as a deformation of the usual (harmonic oscillator)
commutation relations, with ${\cal N} = 2p$ being the deformation
parameter. The form of (\ref{qqanh}) and (\ref{anh}) indicates
that for the low lying levels of the Hamiltonian (\ref{hamq}) the spectrum
corresponds to (\ref{vibr}), the Morse eigenvalues. More
generally, one may consider the parametrization (\ref{qqanh}) to mean
that, up to order $1/p$, the $HW_q$ algebra contracts to
$SU(2)$. Although the range covered by the full Morse eigenvalue
(\ref{anh}) is not consistent with the expansion (\ref{qqanh}),
the approximation is useful nevertheless to provide a physical
interpretation for $p$ or $q$ in terms of the Morse anharmonicity
parameter [2].

For $q\leq 1$, we thus retrieve the case of low-energy harmonic
and anharmonic vibrations in molecules and solids. The quantum
parameter $q$ (or the related parameter $p$ ) would naturally
appear in all vibrational related properties such as infrared and
Raman spectroscopy, partition functions, specific heat, thermal
expansions \cite{lemus:chem, angelova:trjphys, angelova:pub,
cooper:physrev, gupta:modphys}. The
case of \ $q>1 $ is also very interesting as it is related to
Bose-Einstein condensation and superfluidity \cite{marko:physlet,
monteiro:physrevlet}. Under this conditions one may define a new
parameter\ $S\, \equiv\, 1/(q-1)$\ and the approximation
equivalent to (\ref{qqanh}), leads to the noncompact $SU(1,1)$ algebra.
This case is currently under investigation.

\bigskip \bigskip

\section{\protect\bigskip $q$-Bosons at Roots of Unity}

\subsection{Finite-dimensional systems}

In the previous section we considered the system given by
(\ref{qanh}) for real values of the deformation parameter. But it
may be very interesting also in the case when the deformation
parameter $q$ is a root of unity. The reason is that in that
case we can also define finite-dimensional representations. Note
that this case is not related to the bosonic relations (\ref{anh}).

Before proceeding further, we consider a restriction in the
basis of the algebra, namely, we may use only the operator
$K\equiv q^{\hat{n}}$ but not $
\hat{n}$, the reason being that the Casimir operator depends only on
$q^{
\hat{n}}$. The commutation relations are then:
\begin{equation}
\lbrack a,a^{\dagger }]=K\ ,\quad K\,a=q^{-1}\,a\,K\ ,\quad
K\,a^{\dagger }=q\,a^{\dagger }\,K.
\end{equation}

We start with the generic case for the deformation parameter.
Let $|\rangle \,$ be the vacuum which is annihilated by the
operators lowering the boson number and is an eigenvector of the
number operator:
\begin{equation}
a\,|0\rangle \,=0\ ,\qquad K\,|\rangle \,=\, q^\nu |0\rangle\, ,
\qquad ({\rm or}\ \ \hat{n}|\rangle \,=\nu |0\rangle) \,
\end{equation}
where $\nu $ for the moment is an arbitrary complex number. The
states of the system are built by applying the operators raising
the boson number:
\begin{equation}
|k\rangle \ \equiv \ (a^{\dagger })^{k}\,|0\rangle \,.
\end{equation}
The action of the algebra on the basis $|k\rangle \,$ is:
\begin{eqnarray}
&&K|k\rangle \,=q^{\nu +k}|k\rangle \, \label{action} \nonumber
\\ &&a\,|k\rangle \,=q^{\nu }\,\frac{q^{k}-1}{q-1}\,|k-1\rangle
\,
\nonumber \\
&&a^{\dagger }\,|k\rangle \,=|k+1\rangle .
\end{eqnarray}
We denote this representation space by $V_{\nu}\,$.
Clearly, for generic deformation parameter $q$, $V_{\nu}$ is
infinite-dimensional.

The only way to have a finite-dimensional representation is to
suppose that $ q$ is a nontrivial root of unity, {\textit i.e.},
$q^{N}=1$ for some natural number $N>1$. In this case we have:
\begin{equation}
a\,|N\rangle \,=q^{\nu }\,\frac{q^{N}-1}{q-1}\,|N-1\rangle \,=0.
\label{trunk}
\end{equation}
Then all states $\ |k\rangle \,\ $ with $k\geq N$ form an
infinite dimensional invariant subspace, which we denote by
$I_{\nu }$. Indeed, $K$ is diagonal, $a^{\dagger }$ raises the
boson number of $|k\rangle \,$, and the lowering operator $a$ has
$|N\rangle \,$ as a boundary for its action, since it annihilates
this state. Now, we obtain a finite-dimensional representation
space as the following factor-space:
\begin{equation} \label{factor}
F_\nu \ =\ V_\nu/ I_\nu .
\end{equation}
The space $F_\nu$ is $N$-dimensional.

Another way to obtain a finite-dimensional space is by noting
that the above structure is periodic. To see this we consider the
states $|k+mN\rangle $ for fixed $k<N$ and for all non-negative
integer $m$. Then we have:
\begin{eqnarray}
&&K|k+mN\rangle \,=q^{\nu +k+mN}|k\rangle \,=q^{\nu +k}|k\rangle
\,
\label{paction} \nonumber \\
&&a\,|k+mN\rangle \,=q^{\nu
}\frac{q^{k+mN}-1}{q-1}\,|k+mN-1\rangle
\,=q^{\nu }\frac{q^{k}-1}{q-1}\,|k+mN-1\rangle \, \nonumber \\
&&a^{\dagger }\,|k+mN\rangle \,=|k+mN+1\rangle \,
\end{eqnarray}
{\textit i.e.}, the action of the algebra on all states
$|k+mN\rangle $ (for fixed $k$) coincides.

Thus, we can identify these states between themselves and it is
enough to consider the states $|k\rangle $ with $k<N$. Let us
denote these identified states by $\widetilde{|k\rangle }$,
$k=0,\dots ,N-1$. They form a finite-dimensional representation
space $\tilde F_{\nu }$ which has the same dimension as $F_{\nu
}$. The action of the algebra on these states is:
\begin{eqnarray}
&&K\,\widetilde{|k\rangle }=q^{\nu }\widetilde{|k\rangle }
\label{faction}
\nonumber \\
&&a\,\widetilde{|k\rangle }=q^{\nu
}\frac{q^{k}-1}{q-1}\,\widetilde{ |k-1\rangle } \nonumber \\
&&a^{\dagger }\,\widetilde{|k\rangle }=\widetilde{|k+1\rangle }\
,\qquad k<N-1 \nonumber \\ &&a^{\dagger
}\,\widetilde{|N-1\rangle }=\widetilde{|0\rangle }\
\label{alg}
\end{eqnarray}
where in the last line we have used the identification of
$\widetilde{ |N\rangle }$ with $\widetilde{|0\rangle }$.

We have thus obtained an interesting finite-dimensional system.
In this system the boson number lowering operator acts in the
usual way, (in particular, it annihilates the vacuum state
$\widetilde{|0\rangle }$), but the boson raising operator acts
cyclicly. In particular, it has a non-zero action on all states.
Another interesting feature is that the vacuum state may be
obtained not only by the action of the lowering operator but also
by the action of the raising operator producing a jump from
$\widetilde{|N-1\rangle }$ to $\widetilde{|0\rangle }$. One
realization of this operator is a two-level system, obtained for
$N=2$ (equivalent to $q=-1$). For $N>2$, systems with finite
number of levels and population inversion are illustrations of
possible action of these operators.

\subsection{Application to linear lattice with periodic boundary
conditions}

For large $N$ periodicity of the type decribed above appears in
crystals. We will show that it represents the periodic boundary conditions,
first proposed by Born and von
K\'{a}rm\'{a}n \cite{born:physz}. The periodic boundary
conditions are imposed on the translational symmetry, which
strictly speaking is a property of an infinite crystalline
lattice, to allow its use for finite crystals. The boundary
conditions require that every energy eigenfunction $\phi
(\mathbf{r})$ is periodic,
\begin{equation}
\phi (\mathbf{r})=\phi (\mathbf{r}+N\mathbf{t})
\end{equation}
where $\mathbf{r}$ is the position vector, $\mathbf{t}$ is the
vector of primitive translations and $N$ is a large positive
number.

Consider the classic example of a linear lattice of
identical particles with periodic boundary conditions (Fig. 1).
The equilibrium positions of the particles are given by
\begin{equation}
\mathbf{t}_{n}=n\mathbf{t}\mathbf{,\;\;}n=0,1\mathbf{,\ldots ,}N-1
\end{equation}
and the periodic boundary condition requires
\begin{equation}
\mathbf{t}_{N}\equiv \mathbf{t}_{0}\equiv \mathbf{0.}
\end{equation}

\vskip 5mm

$$\vbox{\halign{#\hfil & #\hfil\cr
&\phantom{}$\quad\ \ \rigat{t}$ \cr
&$\ \downb{{0\equiv N}} \ \ \downb{{1}}
\ \ \ \ \downb{{2}}\ \ \ \ \downb{{3}} \ \
 \cdots \ \ \downb{{}}\ \ \ \ \downb{{}}\ \ \ \ \downb{{}}
\ \ \downb{{N-1}} \ $\cr }} $$

\vskip 3mm

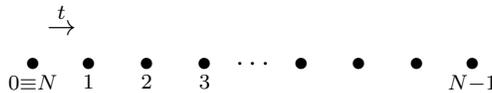
\centerline{Fig. 1. \ {\it The linear lattice}}

\vskip 7mm

The symmetry operations of the linear lattice form a cyclic finite
group of\ order $N $ with a generator, the primitive translation
$\{E|\mathbf{t}\}$.
Here, the Seitz notation is used to represent a translation and \ $E$ is
the identity, $E\equiv \{E|0\}$.

The $n$-th element of the group is
\begin{equation}
\{E|t\}^{n}=\{E|\mathbf{t}_{n}\},\;n=1,2,\ldots ,N-1.
\end{equation}
The product of two elements of the group is an element of the group
\begin{equation}
\{E|t_{m}\}\{E|t_{n}\}=\{E|t_{m+n}\},\;m,n=1,2,\ldots ,N-1, \;
\end{equation}
where $m+n\equiv (m+n)\, ({\rm mod}\,N)$. The identity is
\begin{equation}
\{E|\mathbf{t}\}^{N}\equiv \mathbf{\{}E|\mathbf{t}_{0}\}\equiv \{E|%
\mathbf{0}\}.
\end{equation}

Now, we can show that the generator $\{E|\mathbf{t}\}$ is
isomporphic to the raising operator $a^{\dagger }$. Using the
action (\ref{alg}) of the operator $a^{\dagger }$ on the states
$\widetilde{|k\rangle }$, one can verify that,
\begin{equation}
(a^{\dagger })^{n}\widetilde{|k\rangle }=
\widetilde{|(k+n)\,({\rm mod}\,N)\rangle },\;\;n=0,1,%
\mathbf{\ldots ,}N-1
\end{equation}
and thus:
\begin{equation}
(a^{\dagger })^{m}(a^{\dagger })^{n}
=(a^{\dagger })^{(m+n)\,({\rm mod}\,N)} ,\;\;m,n=0,1,%
\mathbf{\ldots ,}N-1
\end{equation}
Also,
\begin{equation}
(a^{\dagger })^{N}\widetilde{|k\rangle }=(a^{\dagger })^{k}a^{\dagger
}(a^{\dagger })^{N-k-1}\widetilde{|k\rangle }=(a^{\dagger })^{k}a^{\dagger }%
\widetilde{|N-1\rangle }=(a^{\dagger })^{k}\widetilde{|0\rangle }=\widetilde{%
|k\rangle }
\end{equation}
which gives the identity,

\begin{equation}
(a^{\dagger })^{N}=E.
\end{equation}

Thus, the symmetry group of the lattice with periodic boundary conditions is
isomorphic to the finite cyclic group of order $N$ with a
generator, the operator $a^{\dagger }$. This group can be used
with the other symmetry operations of one-dimensional crystalline
or polymer Hamiltonians. To recall, the periodic boundary
conditions \cite{corn:group1, alt:induced} determine the number
of the allowed wave-vector states in the Brillouin zone model and
imply additional selection rules on certain frequencies. The
boundary conditions can be generalised for the three-dimensional
case by introducing raising operators for each dimension.

The deformation at roots of unity, discussed in this section,
imposes in a natural way a periodicity on the boundaries of a finite lattice,
which makes its symmetry compatible with the translational symmetry of the corresponding infinite lattice.

\section{Conclusion}

The application of a $q$-algebra to physical problems is often hampered by a
lack of an appropriate interpretation for the deformation parameters and
often applications are carried out where generalization to $q$-deformed
versions of well known models are made with no simple interpretation.

In this paper we have shown, on the one hand, that a reparametrization of the
deformed algebras (where the classical limit corresponds to $p\to \infty$ in
our example) leads to a natural next order of approximation for the $q$-system.
The $1/p$ approximation of the $HW_q$ example considered in this
paper leads to the $SU(2)$ algebra and to an interpretation of
$p$ in terms of the Morse potential anharmonicity.
Such an approximation may be very useful for the analysis of
other systems in a similar fashion.

On the other hand, we have presented in Section 4 a new
application of the $HW_q$ algebra when $q$ is a root of unity,
which gives a periodic structure described by a
finite-dimensional representation. We have shown that the raising
operator belonging to this structure generates a group isomorphic
to the symmetry group of a linear lattice with periodic boundary
conditions. The latter may provide a useful framework for the
deformation of crystalline or polymer Hamiltonians.

To conclude, we stress that simple approximations and specific realizations
of $q$-algebras, such as the ones discussed in this paper, may shed some
light on the role of these mathematical constructs and open new ways to their
physical interpretation.

\end{document}